\documentclass[proof]{WileyASNA-v1}

\articletype{Article Type}%

\received{2021}
\revised{2021}
\accepted{2021}

\begin{document}

\title{Time-Resolved Photometry of the High-Energy Radiation of M~Dwarfs with the \textit{Star-Planet Activity Research CubeSat (SPARCS)}}

\author[1]{Tahina Ramiaramanantsoa*}

\author[1]{Judd D. Bowman}

\author[1]{Evgenya L. Shkolnik}

\author[1]{R. O. Parke Loyd}

\author[2]{David R. Ardila}

\author[3]{Travis Barman}

\author[2]{Christophe Basset}

\author[4]{Matthew Beasley}

\author[2]{Samuel Cheng}

\author[1]{Johnathan Gamaunt}

\author[2]{Varoujan Gorjian}

\author[1]{Daniel Jacobs}

\author[1]{Logan Jensen}

\author[2]{April Jewell}

\author[5]{Mary Knapp}

\author[6]{Joe Llama}

\author[7]{Victoria Meadows}

\author[2]{Shouleh Nikzad}

\author[8]{Sarah Peacock}

\author[1]{Paul Scowen}

\author[2]{Mark R. Swain}

\authormark{RAMIARAMANANTSOA \textsc{et al}}

\address[1]{\orgdiv{School of Earth and Space Exploration}, \orgname{Arizona State University}, \orgaddress{\state{Arizona}, \country{USA}}}

\address[2]{\orgdiv{Jet Propulsion Laboratory}, \orgname{California Institute of Technology}, \orgaddress{\state{California}, \country{USA}}}

\address[3]{\orgdiv{Lunar and Planetary Laboratory}, \orgname{University of Arizona}, \orgaddress{\state{Arizona}, \country{USA}}}

\address[4]{\orgdiv{Department of Space Studies}, \orgname{Southwest Research Inc.}, \orgaddress{\state{Colorado}, \country{USA}}}

\address[5]{\orgdiv{Haystack Observatory}, \orgname{MIT}, \orgaddress{\state{Massachusetts}, \country{USA}}}

\address[6]{\orgdiv{Science Division}, \orgname{Lowell Observatory}, \orgaddress{\state{Arizona}, \country{USA}}}

\address[7]{\orgdiv{Department of Astronomy}, \orgname{University of Washington}, \orgaddress{\state{Washington}, \country{USA}}}

\address[8]{\orgdiv{Goddard Space Flight Center}, \orgname{NASA}, \orgaddress{\state{Maryland}, \country{USA}}}

\corres{*Tahina Ramiaramanantsoa, ASU/SESE, 781 E. Terrace Mall, Tempe, AZ 85287, USA. \email{tahina@asu.edu}}

\presentaddress{ASU/SESE, 781 E. Terrace Mall, Tempe, AZ 85287, USA.}

\abstract{Know thy star, know thy planet,… especially in the ultraviolet (UV). Over the past decade, that motto has grown from mere wish to necessity in the M~dwarf regime, given that the intense and highly variable UV radiation from these stars is suspected of strongly impacting their planets’ habitability and atmospheric loss. This has led to the development of the \textit{Star-Planet Activity Research CubeSat} (\textit{SPARCS}), a NASA-funded 6U CubeSat observatory fully devoted to the photometric monitoring of the UV flaring of M~dwarfs hosting potentially habitable planets. The \textit{SPARCS} science imaging system uses a 9-cm telescope that feeds two delta-doped UV-optimized CCDs through a dichroic beam splitter, enabling simultaneous monitoring of a target field in the near-UV and far-UV. A dedicated onboard payload processor manages science observations and performs near-real time image processing to sustain an autonomous dynamic exposure control algorithm needed to mitigate pixel saturation during flaring events. The mission is currently half-way into its development phase. We present an overview of the mission’s science drivers and its expected contribution to our understanding of star-planet interactions.  We also present the expected performance of the autonomous dynamic exposure control algorithm, a first-of-its-kind on board a space-based stellar astrophysics observatory.}

\keywords{stars: flare, stars: rotation, ultraviolet: stars, space vehicles: instruments, techniques: photometric}

\jnlcitation{\cname{%
\author{Tahina Ramiaramanantsoa}, 
\author{Judd D. Bowman},
\author{Evgenya Shkolnik},
\author{R. O. Parke Loyd},
\author{David R. Ardila},
\author{Travis Barman},
\author{Christophe Basset},
\author{Matthew Beasley},
\author{Samuel Cheng},
\author{Johnathan Gamaunt},
\author{Varoujan Gorjian},
\author{Daniel Jacobs},
\author{Logan Jensen},
\author{April Jewell},
\author{Mary Knapp},
\author{Joe Llama},
\author{Victoria Meadows},
\author{Shouleh Nikzad},
\author{Sarah Peacock},
\author{Paul Scowen}, and
\author{Mark R. Swain}} (\cyear{2021}), 
\ctitle{Time-Resolved Photometry of the High-Energy Radiation of M~Dwarfs with the \textit{Star-Planet Activity Research CubeSat (SPARCS)}}, \cjournal{Astronomische Nachrichten}.}


\maketitle

\let\thefootnote\relax\footnotetext{\textbf{Abbreviations:} CCD, charge-coupled device; C\&DH, command and data handling; DFT, discrete Fourier transform; EUV, extreme-ultraviolet; FPGA, field-programmable gate array; FUV, far-ultraviolet; \textit{GALEX}, \textit{Galaxy Evolution Explorer}; \textit{HST}, \textit{Hubble Space Telescope}; NUV, near-ultraviolet; PSF, point spread function; \textit{SPARCS}, \textit{Star-Planet Activity Research CubeSat}; UV, ultraviolet.}

\section{Introduction}
\label{sec:Intro}

\begin{figure*}[ht]
	\centerline{\includegraphics[width=16cm]{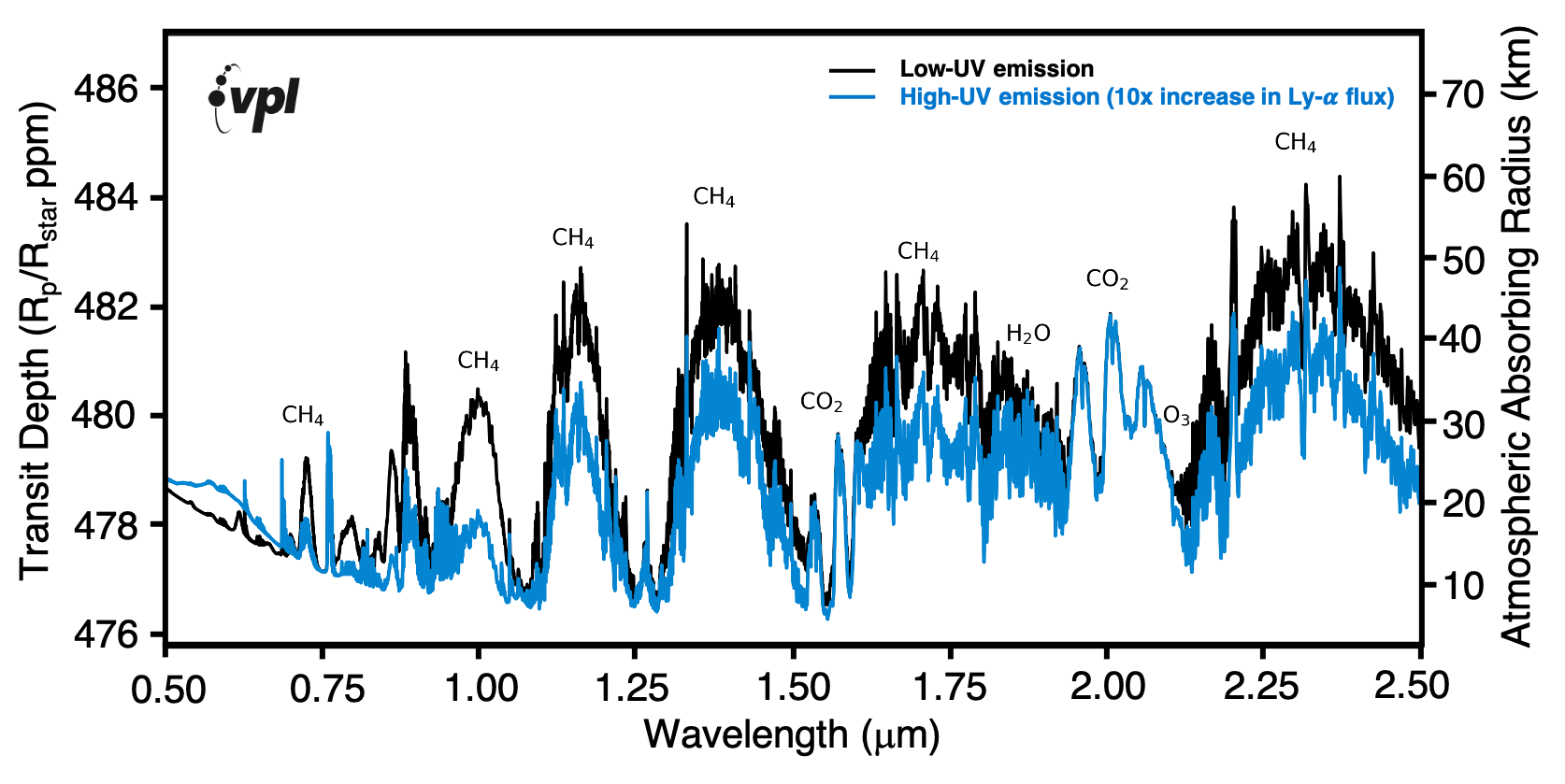}}
	\caption{Simulated transmission spectra of an Earth-like planet in the habitable zone of GJ 436 (M3V). A $10\times$ flux increase in \mbox{Ly-$\alpha$} from the host star strongly enhances photodissociation of methane in the atmosphere of the planet (Davis et al., in prep.).}
	\label{fig:VPL_Transmission_GJ436}
\end{figure*}

Despite their relative faintness and petiteness, M~dwarfs are among the most extreme intrinsic variable stars in the Hertzsprung--Russell diagram. Besides showing low-amplitude rotational modulation arising from corotating photospheric/chromospheric active regions, they also exhibit extreme transient variability associated with erratic flaring events. Recent multi-wavelength observations of M~dwarfs revealed that the flux increase during a flare event can be more extreme in the ultraviolet (UV) than in the optical, e.g. a flare that peaks at $\sim$$0.01\times$ the star's quiescent flux in the optical can have a UV counterpart that brightens by a factor of $\sim$$14000$ \citep{MacGregor2021ApJ...911L..25M}. Moreover, UV superflare events (emitting energy $\geq$$10^{33}$~erg) leading to $\geq$$200\times$ flux increases are expected to occur on a daily basis for young, active M~dwarfs \citep{Loyd2018ApJ...867...70L}. This becomes increasingly a matter of concern in studies of habitability around M~dwarfs, as theoretical investigations indicate that M~dwarf UV radiation variability may strongly affect the atmospheric composition of their planets, be it through photodissociation of molecules by \mbox{far-UV} (FUV) and \mbox{near-UV} (NUV) radiation, or through atmospheric heating and escape by \mbox{extreme-UV} (EUV) photons \citep[e.g.][]{Segura2010AsBio..10..751S,Hu2012ApJ...761..166H,Owen2012MNRAS.425.2931O,Luger2015AsBio..15..119L}. Figure~\ref{fig:VPL_Transmission_GJ436} shows that a $10\times$ flux increase in \mbox{Ly-$\alpha$} for an M3~dwarf can lead to strong depletion of methane in the atmosphere of a modern Earth-like planet in the habitable zone of the star. Although a $10\times$ \mbox{Ly-$\alpha$} flux increase would be relatively extreme, it is not completely unreasonable to imagine such a situation given that the Proxima Cen flare that increased the star's FUV continuum flux by a factor of $\sim$$14000$ made a $5\times$ \mbox{Ly-$\alpha$} flux increase (\mbox{R. O. Parke Loyd}, private communication). Furthermore, there are currently not enough observations of M~dwarf UV flares to constrain these theoretical studies. The longest intensive UV monitoring of an M~dwarf was done thus far with the \textit{Hubble Space Telescope} (\textit{HST}) and was limited to $\sim$$30$~h \citep{MacGregor2021ApJ...911L..25M}. These circumstances led to the development of the \textit{Star-Planet Activity Research CubeSat} \citep[\textit{SPARCS;}][]{Shkolnik2016apra.prop...98S}, as space-based SmallSat mission designed to address the need for a comprehensive picture of M~dwarf activity in the UV.

\section{UV photometric monitoring of M~dwarfs with \textit{SPARCS}}
\label{sec:SPARCS}

\subsection{The mission}
\label{subsec:SPARCS_Overview}

\begin{figure}[t]
	\centerline{\includegraphics[width=86mm]{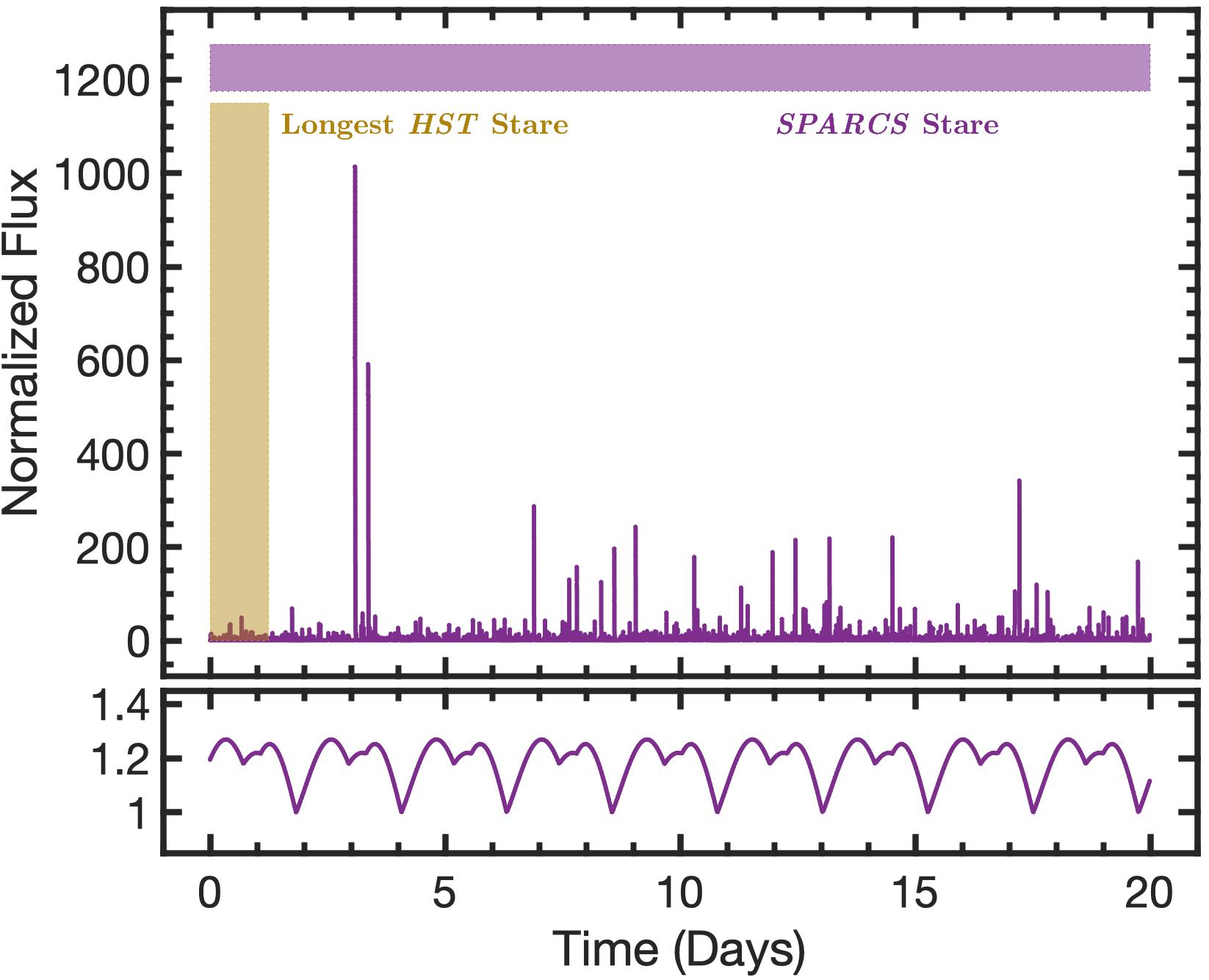}}
	\caption{Simulated $20$-day noise-free and gap-free light curve of the active M4V star AD Leo in the \textit{SPARCS} FUV passband, with the bottom panel zooming on the out-of-flare light curve exhibiting low-amplitude rotational modulation arising from two bright chromospheric active regions located near the stellar equator ($\theta=90^{\circ}$ and $\theta=80^{\circ}$), nearly antipodal to each other ($\Delta\varphi = 210^{\circ}$), and corotating with the star at $P_{rot}=2.2399$~d. \label{fig:sim_lc_ADLeo}}
\end{figure}

\textit{SPARCS} is a NASA-funded 6U ($30$~cm~$\times$~$20$~cm~$\times$~$10$~cm) CubeSat observatory (PI Shkolnik) under development that hosts a 9-cm f/6 reflective telescope and a dual-band UV camera (SPARCam) to perform long-term, high-cadence photometric monitoring of M~dwarf flaring and chromospheric activity in the NUV ($258-308$~nm) and the FUV ($153-171$~nm). Expected to be ready for insertion into a heliosynchronous orbit in 2023, the spacecraft will monitor a sample of twenty young and old M~dwarfs ($10$~Myr -- $5$~Gyr) over an expected mission lifetime of $1$~yr. A heliosynchronous orbit allows for decent thermal stability and optimized continuity in target monitoring. The \textit{SPARCS} dual-band, high-cadence, long time baseline observations (Figure~\ref{fig:sim_lc_ADLeo}) will enable measurements of M~dwarf UV flare color, energies, occurrence rate, and duration of quiescent and flaring states for both active and inactive M~dwarfs. It is also anticipated that \textit{SPARCS} observations will help predict M~dwarf EUV flux to better than a factor of two, and will allow for the development of improved EUV--NUV M~dwarf upper-atmosphere model spectra \citep[see][and references therein]{Peacock2020ApJ...895....5P} that will serve as more reliable inputs to the modelling of the atmospheres of planets around M~dwarfs.

Additionally, \textit{SPARCS} has a relatively wide field-of-view of $40^\prime$ that will enable ancillary science on secondary targets (e.g. FGKM stars, AGNs) in the vicinity of the primary target M dwarfs. \textit{SPARCS} final data products, including raw images, processed images, and light curves of science targets, will be ultimately archived at the Mikulski Archive for Space Telescopes.

\subsection{Science payload operations}
\label{subsec:SPARCS_Payload_Ops}

\textit{SPARCS} onboard payload operations are controlled by a custom Python-based software running on a dedicated payload processor. Payload operations can be divided into three main categories: communications with the spacecraft's command and data handling (C\&DH) computer, control and monitoring of the temperatures of the SPARCam detectors, and science observations.

\paragraph{\emph{Payload-to-spacecraft communications}}

Payload commands transmitted from the ground to the spacecraft are first received by the spacecraft's C\&DH computer, which then transfers them to the payload processor. Hence the latter is tasked to promptly detect new incoming payload commands, reassemble them, and distribute them into the filesystem. The payload processor is also in charge of transferring payload engineering data (time series of detector temperature measurements) and science data (raw science images and calibration images) to the C\&DH, which will relay them to the ground during downlink operations.

\paragraph{\emph{Active detector thermal control}}

SPARCam uses two back-illuminated, delta-doped, frame transfer charge-coupled devices (CCDs). During routine science observations, the CCDs have to be maintained at \mbox{$(-35\pm3)^\circ$C} to keep dark current noise in the detectors as low as $\sim$$0.1$~electrons/pixel/s. The payload processor achieves that by commanding a temperature controller board to perform proportional-integral-derivative temperature control and regular temperature measurements at a specified cadence.  

\paragraph{\emph{Science observations}}

Lastly, the payload processor manages science observations, which consists of controlling SPARCam exposures, retrieving images from it, processing images, extracting small image subrasters containing science targets from the raw full-frame image, and saving them (or the raw full-frame image when requested) to disk. Near real-time onboard image processing is required in order to support source finding and autonomous detector exposure time and gain control.

\subsection{Routine science observations with onboard autonomous exposure control}
\label{subsec:SPARCS_Science_Obs}

\begin{figure*}[ht]
\centerline{\includegraphics[width=16cm]{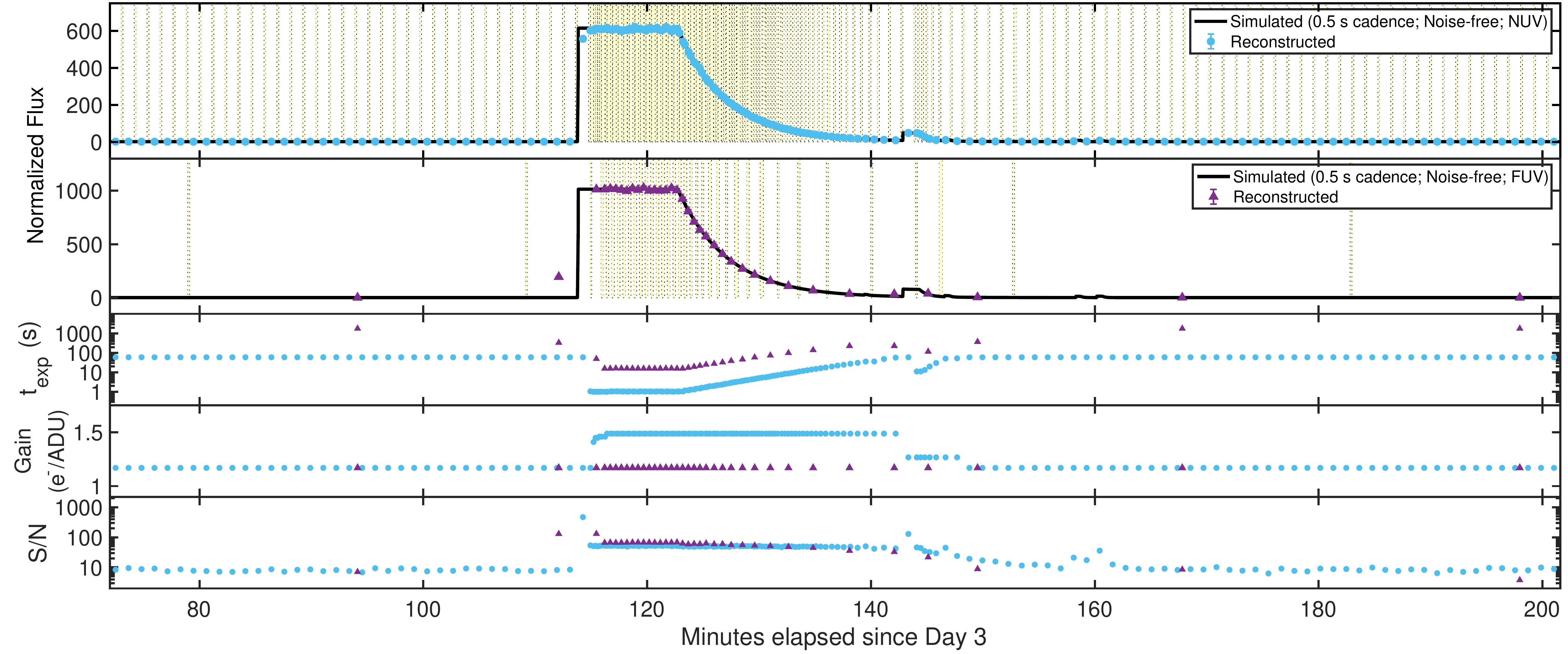}}
\caption{Excerpt of a simulated \mbox{$20$-day} \textit{SPARCS} observing run on AD Leo, used to test the robustness of the segment of the payload software that executes science observations with autonomous exposure control. Shaded regions demarcate overheads. The algorithm shows satisfactory response to the steep flare rise in the simplified temporal flare shape, implying that it will also react well on real M~dwarf UV flares which are expected to have more gradual rises.}
\label{fig:sim_lc_ADLeo_Flares}
\end{figure*}

Previous space-based astrophysics observatories devoted to long-term photometric monitoring of stars (e.g. \textit{MOST}, \textit{CoRoT}, \textit{Kepler}, \textit{BRITE}, \textit{TESS}) performed target field monitoring at constant detector exposure times and gains. The \textit{SPARCS} payload software also has that basic capability. However, unlike its optical predecessors, \textit{SPARCS} adopts an onboard autonomous exposure control as the default observing mode in order to mitigate the occurence of pixel saturation when observing strong M~dwarf UV flaring events, which have far weaker optical counterparts.

\subsubsection{Onboard image processing and autonomous exposure control}
\label{subsubsec:SPARCS_Science_Obs_AEC_proc}

The \textit{SPARCS} onboard dynamic exposure control has to react quickly at the start of a flaring event and use minimal computer resources. Given that the bias-subtracted maximum of the science target’s point spread function (PSF) is directly proportional to the exposure time and inversely proportional to gain, an algorithm analogous to a proportional controller is sufficient. In a freshly-acquired image, the bias-subtracted maximum of the primary target source’s PSF is measured and compared to a setpoint value that has been specified for that parameter. If the measured value is lower (resp. higher) than the setpoint, the exposure time is increased (resp. decreased) by the appropriate multiplicative factor. The gain is changed after the exposure time reaches its minimum allowable value. However, if saturation is detected, the exposure time and the gain are respectively set to their minimum and maximum values. Conversely, when the primary science target is not detected, the imaging system is immediately set to its highest sensitivity configuration, i.e. maximum allowable exposure times and minimum allowable gains. The current version of the software is configured such that detection of a flaring event in the NUV channel will trigger an abort of the ongoing FUV exposure and new image acquisitions at reduced exposure times (or increased gains) in both channels. The NUV channel is chosen to be the trigger of exposure abort upon flare detection because \textit{SPARCS} targets are brighter in the NUV such that their required quiescent exposure times are much shorter in the NUV than in the FUV and, therefore flares are more often detected first in the NUV exposures than in the FUV ones.

To support source location and autonomous exposure control, images are subject to minimal processing: bias subtraction, flat-fielding, dark correction, as well as bad pixel and cosmic ray hit corrections. Source location is done with \textsc{Source Extractor} \citep{Bertin1996A&AS..117..393B}. 

\subsubsection{Simulated observing runs}
\label{subsubsec:SPARCS_Science_Obs_tests}

Simulated observing runs were conducted to test the robustness of the segment of the \textit{SPARCS} payload software that regulates science observations.

\begin{figure*}[t]
\centerline{\includegraphics[width=16cm]{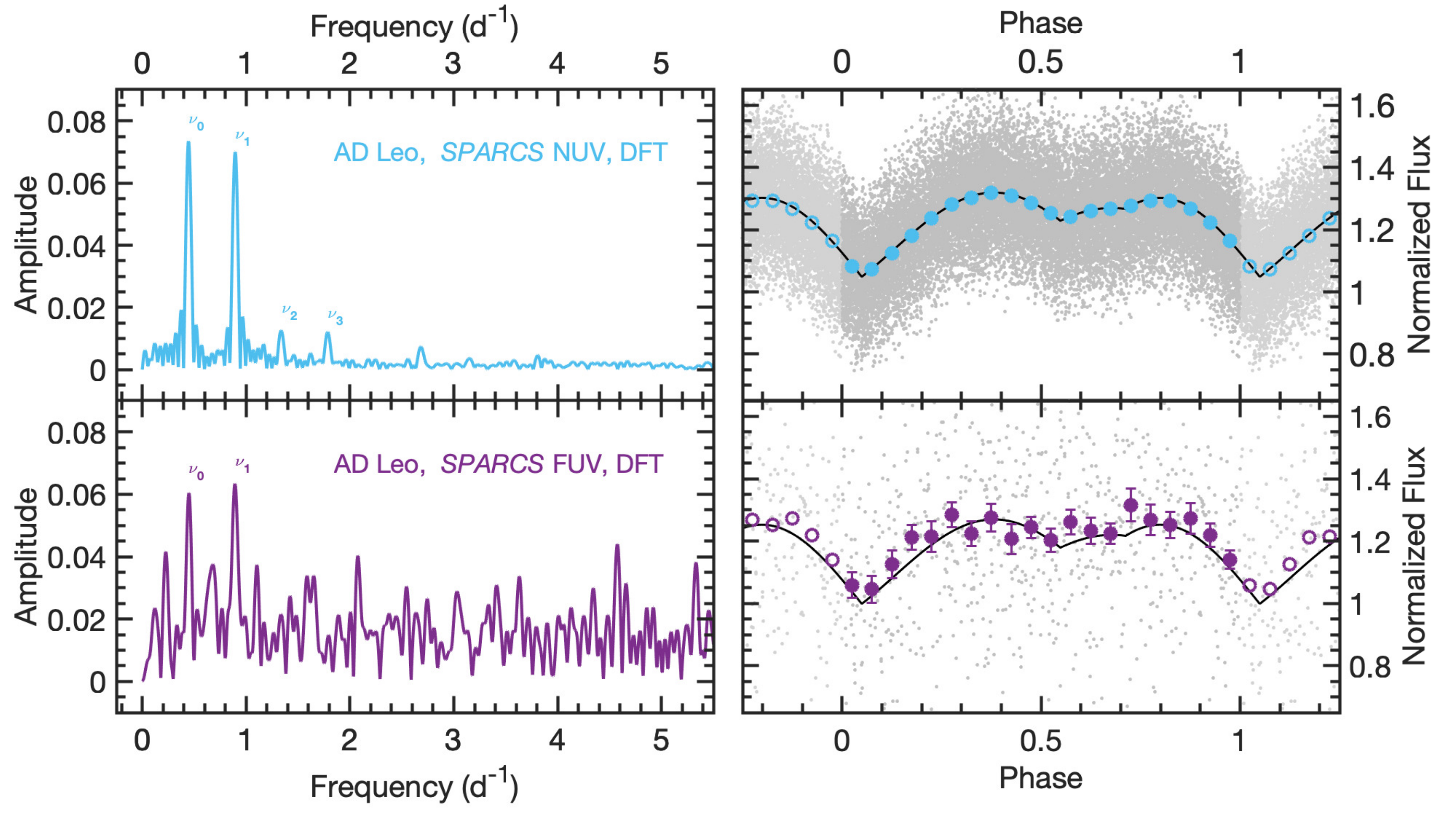}}
\caption{A $20$-day monitoring of AD Leo with \textit{SPARCS} would reveal rotational modulation that would look like this if the star had two dominating bright active regions nearly antipodal to each other close to the stellar equator. \textit{Left:} Amplitude spectra from the discrete Fourier transforms (DFTs) of the out-of-flare light curves. \textit{Right:} The out-of-flare light curves phase-folded on the stellar rotation period $P_{rot}=2.2399$~d. Small grey points are the original observations, while larger points are averages over $0.05$ phase bins. The thin continuous lines are the original (noise-free) out-of-flare synthetic light-curves.}
\label{fig:sim_lc_ADLeo_Rotation}
\end{figure*}

Noise-free, high-cadence ($0.5$~s) time series of stellar flux measurements were created for representative target M~dwarfs such as AD Leo in the \textit{SPARCS} bandpasses (Figure~\ref{fig:sim_lc_ADLeo}). The model light curves incorporate flaring events and rotational modulation. Flares are generated with the tool described in \citet{Loyd2018ApJ...867...71L}, relying on empirically-based M~dwarf UV flare frequency distributions but assuming a flare temporal profile approximated as a combination of a step-wise rising phase, a plateau, and an exponential decay phase \citep[see also Figure~20 in][]{Loyd2018ApJ...867...71L}. That modeling approach allows us to check how the onboard exposure control algorithm responds to the worst case scenario of the steepest possible flare rising phase while adopting realistic M~dwarf UV flare amplitudes and durations. If the control algorithm reacts well to such step-wise flare rises, it is expected that it will respond even better to real M-dwarf UV flares, which will have more gradual rising phases, one or more sharp peaks in lieu of a plateau, and variable decay rates \citep{Loyd2014ApJS..211....9L}. 
Additionally, based on limited observational information on rotational modulation in M~dwarfs in the UV \citep{Miles2017AJ....154...67M,dosSantos2019A&A...629A..47D}, rotational modulation with an amplitude of $\sim$$25\%$ induced by bright chromospheric active regions is injected in the synthetic noise-free light curves (Figure~\ref{fig:sim_lc_ADLeo}) using an analytical spot modelling formalism \citep{Lanza2003A&A...403.1135L}.

To simulate an observing run, whenever an image acquisition with a given exposure time is triggered in any channel, the average flux of the target of interest at mid-exposure is evaluated from the noise-free one-dimensional reference light curve. Subsequently, the average flux is converted into photo-electron counts at the focal plane by taking into account the predicted various attenuation factors through the \textit{SPARCS} imaging system. Photo-electron counts are divided by the detector gain (electrons/ADU) and spread over a Gaussian PSF in a synthetic full-frame image. Other sources brighter than $20^{\rm th}$ magnitude in the \textit{GALEX} (\textit{Galaxy Evolution Explorer}) NUV passband and within a radius of $20^\prime$ of the primary target are added to the synthetic image but considered non-variable. Sky background, dark current noise, readout noise, a bias offset of 200 ADU, and cosmic ray hits are also added to the simulated image.

Figure~\ref{fig:sim_lc_ADLeo_Flares} depicts a segment of a $20$-day simulated observing run on AD Leo. The exposure control algorithm shows very good response to step-wise flare rises, and therefore it is expected to respond well to real M-dwarf UV flares that will have less steep rising phases. Overheads of $\sim$$9.7-19.4$~s occur between two consecutive image integrations in any one channel due to image transfer from the camera to the payload processor, image assembling, image processing, exposure control, and image writing onto disk.

Analysis of the out-of-flare light curves obtained from the $20$-day simulated observing run on AD Leo is summarized in Figure~\ref{fig:sim_lc_ADLeo_Rotation}. For each channel, the out-of-flare light curve was extracted by first keeping only observations taken at the maximum exposure time ($1$~min in the NUV and $30$~min in the FUV in this example), then performing iterative sigma clipping. Amplitude spectra from the discrete Fourier transforms (DFTs) of the out-of-flare observations were evaluated with \textsc{Period04} \citep{LenzBreger2005CoAst.146...53L} up to the Nyquist frequency of $\sim$$629$~d$^{-1}$ for the NUV observations and $\sim$$24$~d$^{-1}$ for the FUV data. In general, a frequency peak in the amplitude spectrum is considered statistically significant only at S/N$\geq$$4$ \citep{Breger1993A&A...271..482B,Kuschnig1997A&A...328..544K}. The NUV amplitude spectrum exhibits statistically significant peaks at the rotation frequency and its first harmonic (both with S/N$\sim$$6.0$), as well as minor peaks at the second and third harmonics (S/N$\sim$$2.9$ and $\sim$$2.8$, respectively), while only peaks at the rotation frequency and its first harmonic barely stand out (S/N$\sim$$2.9$ and $\sim$$3.2$, respectively) in the amplitude spectrum of the FUV observations which contains $\sim$$30$~times fewer data samples than the NUV channel. The rotation curves are relatively well-reconstructed in both channels. This also illustrates the power of simultaneous observations in confirming signals that could be deemed not statistically significant in the amplitude spectrum of the more sparsely-sampled data set. Such observations are crucial for stars with less well-constrained rotation periods.

\section{Conclusion}
\label{sec:Conclusion}

Currently half-way into its development phase, the \textit{Star-Planet Activity Research CubeSat (SPARCS)} mission is a SmallSat observatory devoted to the long time baseline, high-cadence, simultaneous FUV and NUV photometric monitoring of the flaring and chromospheric activity of M~dwarfs. The long time baseline and high-cadence aspects of \textit{SPARCS} monitoring campaigns are expected to enable observations of M~dwarf UV flares across a much broader range of energies than previous short-term UV monitoring efforts have achieved, and therefore expected to significantly improve M~dwarf UV flare frequency distributions. The \textit{SPARCS} dual-band UV observations will also enable the prediction of M~dwarf EUV flux to better than a factor of two, and provide better insights on the effects of M~dwarf UV radiation on the atmospheres of \mbox{their planets.}

\textit{SPARCS} will monitor M~dwarfs using a $9$-cm telescope and two back-illuminated, delta-doped, UV-optimized CCDs. Onboard science operations are managed by a custom fully Python-based software running on a dedicated payload processor. The payload software is able to run monitoring campaigns at constant detector exposure time and gain, but due to the expected high amplitudes of M~dwarf UV flares, observations throughout the nominal mission will be conducted using a feature of the software that autonomously adjusts detector exposure times and gains to mitigate the occurence of pixel saturation during observations of flaring events. \textit{SPARCS} will be the first space-based stellar astrophysics observatory that adopts such an onboard autonomous exposure control. The performance of the control algorithm and the science observation segment of the software was tested using simulated observing runs involving empirically-constrained model light curves as well as a full-frame image simulator that incorporates the predicted properties of the \textit{SPARCS} imaging system. The control algorithm reacts very well to the worst-case scenario of step-wise flare rises, which means it will also respond well --- if not better --- to the more gradual rises in actual M~dwarf UV flares. 
The analysis of the out-of-flare light curves generated by the simulated observing runs suggests that \textit{SPARCS} will also allow for the study of low-amplitude rotational modulation in relatively bright targets.  It also illustrates the importance of contemporaneous observations for efficiently confirming the astrophysical origin of signals in sparsely-sampled data, which may be particularly useful when the stellar rotation is poorly constrained or unknown.


\section*{Acknowledgments} 

The \textit{SPARCS} team acknowledges support from the  \fundingAgency{NASA} Astrophysics Research and Analysis program (\fundingNumber{NNH16ZDA001N-APRA; 80NSSC18K0545}). A portion of the research was carried out at the Jet~Propulsion Laboratory, California Institute of Technology, under a contract with \fundingAgency{NASA}  (\fundingNumber{80NM0018D0004}).

\bibliography{SPARCS_XMMNewton2021}%
\bibliographystyle{Wiley-ASNA}

\end{document}